\documentclass{optica-article}

\journal{opticajournal} % for journals or Optica Open

\articletype{Research Article}

\usepackage{lineno}
\begin{document}

\title{Inherently unpredictable beam steering for quantum LiDAR}

\author{Junyeop Kim,\authormark{1,$\dag$} Dongjin Lee,\authormark{1,2,$\dag$} Woncheol Shin,\authormark{1,2} Yeoulheon Seong,\authormark{1} and Heedeuk Shin\authormark{1,*}}

\address{\authormark{1}Department of Physics, Pohang University of Science and Technology (POSTECH), Pohang, 37673, Korea\\
\authormark{2}Quantum Technology Research Laboratory, Electronics and Telecommunications Research Institute (ETRI), 218 Gajeong-ro, Yuseong-gu, Daejeon 34129, Republic of Korea\\
\authormark{$\dag$}The authors contributed equally to this work.}

\email{\authormark{*}heedeukshin@postech.ac.kr} %% email address is required; see note below about the corresponding author designation

% use {asbstract*} to suppress the copyright line. Copyright information will be added in production

\begin{abstract*} 
Quantum LiDAR offers noise resilience and stealth observation capabilities in low-light conditions. In prior demonstrations, the telescope pointing was raster-scanned, making the observation direction predictable from the pointing direction. However, while Quantum LiDAR can enable stealth observation, operational stealth is enhanced by inherently unpredictable beam steering. Here, we introduce a novel stealth beam steering method that is fundamentally immune to prediction. In a photon pair, the probe photon undergoes diffraction in an unpredictable direction at a grating due to wavelength randomness. The arrival time of the heralding photon, delayed by propagation through a dispersive medium, enables the determination of the probe photon’s diffraction direction. Our method successfully detects multiple targets in parallel, demonstrating up to a 1000-fold enhancement in signal-to-noise ratio compared to classical LiDAR. This breakthrough establishes a new paradigm for quantum-enhanced sensing, with far-reaching implications for quantum metrology, secure communications, and beyond. 
\end{abstract*}

\section{Introduction}\label{sec1}

Quantum information technologies, leveraging the quantum states of light, can surpass classical optical limitations but require long coherence times and low loss. However, quantum sensing fields, including quantum ghost imaging~\cite{pittman1995optical}, quantum illumination~\cite{lloyd2008enhanced, tan2008quantum, shapiro2009quantum, shapiro2020quantum}, and quantum light detection and ranging (quantum LiDAR)~\cite{england2019quantum}, utilize the photon-pair correlation to distinguish desired signal photons from background noise, significantly enhancing the signal-to-noise ratio (SNR) in low-light environments, with potential applications in remote sensing on harsh environments. 

Quantum enhancement under phase-sensitive joint measurement surpasses optimal classical limits and has been experimentally validated in various fields requiring higher distinguishability~\cite{reichert2022quantum, zhuang2022ultimate, huang2021quantum, shapiro2009defeating, zhang2013entanglement, shapiro2014secure, zhang2015entanglement}. Moreover, it has been suggested that simultaneous measurement of a target's range and velocity at the Heisenberg limit can be realized by utilizing displaced squeezed light and homodyne detection~\cite{reichert2024heisenberg}. There are approaches in other directions, such as phase-insensitive coincidence counting measurements that exploit the strong correlation between photon pairs. In this case, the classical counterpart is a probe intensity equivalent classical photon counting measurement system~\cite{staffas2024frequency}. The corresponding enhancement factor, the SNR enhancement beyond classical systems, is proportional to the correlation strength and coincidence-to-accidental ratio (CAR) between photon pairs~\cite{lopaeva2013experimental, liu2019enhancing, liu2020target, murchie2024object}. As CAR can be enhanced by reducing the pair generation rate per mode~\cite{shin2023photon, wang2014multichannel}, increasing mode separation in the temporal, spatial, and spectral domains can be beneficial for quantum imaging and LiDAR. A parallel observation strategy, similar to that used in quantum ghost imaging, reduces the photon-pair generation rate per spatial mode~\cite{klyshko1988combine, meyers2008ghost}, yielding large CAR values per spatial mode. 

In quantum LiDAR, implementing a parallel observation strategy is significantly challenging because it requires capturing spatial information from multiple pixels in a single measurement by detecting randomly scattered photons. To date, quantum illumination and quantum LiDAR have primarily been implemented by scanning a single spatial-mode beam, employing a high time-resolution detector for efficient detection in low-light environments~\cite{torrome2024advances}. However, conventional beam steering methods, including mechanical and electro-optic scanning~\cite{meyer1972optical, carlson1988electronic, goyer1963laser,Bi2021}, inherently allow prediction of the beam's propagation direction through mirror orientation or electronic signals, even though the photons themselves are not visible, making it difficult to achieve ideal stealth observation. Furthermore, parallel observation — as seen in quantum ghost imaging~\cite{gregory2020imaging}— yields its inherent spatial unpredictability, meaning the position of one photon from a generated photon pair remains unknown until its twin photon's position is measured. However, quantum LiDAR with parallel observation requires special sensors satisfying both spatially resolved detectors and high temporal resolution ($<$ 1 ns), which presents a significant technical challenge~\cite{EMCCD}. Consequently, a novel technical approach is necessary to preserve directional randomness while ensuring precise directional measurement.

In this paper, we propose a novel approach, referred to as quantum-enhanced parallel light detection and ranging (QEP-LiDAR), leveraging the inherent unpredictability originating from quantum fluctuations, creating a secure and efficient detection mechanism. From photon pairs exhibiting strong temporal-spectral correlations, signal photons are reflected from objects, while idler photons are used to determine the wavelength information. This method ensures that the observation direction remains unknown until the idler photon measurement is conducted, significantly enhancing the system’s stealth capabilities. Furthermore, the system achieved a significantly superior SNR in noisy conditions than classical LiDAR. This breakthrough will not only advance quantum sensing but also open new avenues for stealth and secure detection applications.

\section{Results}\label{sec2}

\subsection{Parallel target detection with temporal-spectral correlated photon pairs}\label{subsec1}

\begin{figure}[htbp]
\centering\includegraphics[width=7cm]{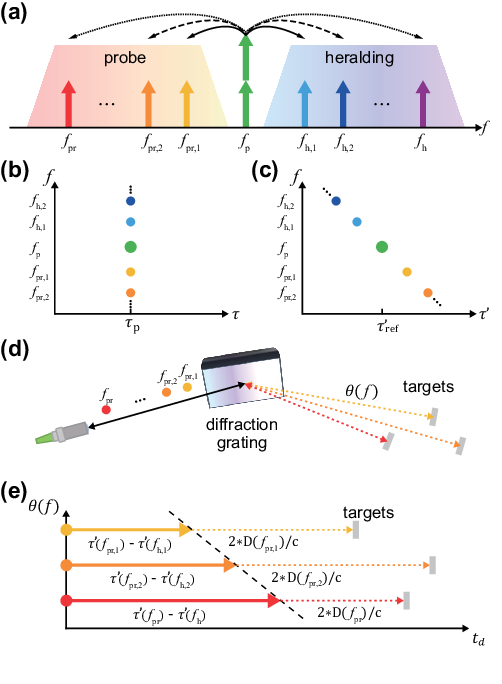}
\caption{Schematic diagrams of quantum-enhanced parallel light detection and ranging (QEP-LiDAR). (a) Spectral diagram of spontaneous four-wave mixing (SpFWM), where $f_{\mathrm{p}}$ denote the frequency of pump photons, and $f_{\mathrm{pr, i}}$ and $f_{\mathrm{h, i}}$ (for $\mathrm{i}=1,2,\ldots$) represent the frequencies of signal (probe) and idler (heralding) photons, respectively. (b) Temporal distribution of SpFWM-generated photons relative to the pump time ($\tau_{\mathrm{p}}$). (c) Temporal separation of photons after passing through a dispersive medium with a constant group velocity dispersion. $\tau^\prime_{\mathrm{ref}}$ is the reference time for the pump light to pass through the dispersion medium. (d) Spatial separation of probe photons by a diffraction grating, where $\theta \left( f \right)$ indicates the direction of each photon as a function of its frequency. (e) Schematic of distance measurement to targets using the QEP-LiDAR method, where $t_d$ is the temporal delay between heralding and probe photons.} \label{fig1}
\end{figure}

This section explains the working principles of QEP-LiDAR, which utilizes the strong temporal-spectral correlations between paired photons. Fig. \ref{fig1} illustrates the system's schematic diagrams. Photon pairs are generated through spontaneous four-wave mixing (SpFWM), as shown in Fig. \ref{fig1}(a), where two pump photons ($f_{\mathrm{p}}$) are converted into a probe photon ($f_{\mathrm{pr}}$) and a heralding photon ($f_{\mathrm{h}}$). This process occurs at the generation time ($\tau_{\mathrm{p}}$) under the energy conservation condition ($2f_{\mathrm{p}}=f_{\mathrm{h}}+f_{\mathrm{pr}}$), as shown in Fig. \ref{fig1}(b). Notably, since the SpFWM process originates from vacuum fluctuations, the frequency of the probe photon remains unpredictable until measurement, forming the basis of the inherent unpredictability in this study.

If the photons pass through a dispersive medium with constant anomalous group velocity dispersion ($\beta_2<0$), their arrival times ($\tau^\prime \left( f \right)$) become frequency-dependent, as shown in Fig. \ref{fig1}(c), and  can be described by the following equation:
\begin{align}
\tau^\prime\left( f \right) &= \tau_{\mathrm{p}}+L\cdot f \cdot \beta_2 = \tau^\prime_{ref}+L\cdot \left( f-f_{\mathrm{p}}\right) \cdot \beta_2, \label{eq1}
\end{align}
where $\tau^\prime_{\mathrm{ref}} \equiv \tau_{\mathrm{p}}+L \cdot f_{\mathrm{p}}\cdot \beta_2$ is the reference time, which represents the pump's arrival time through the dispersion medium and $L$ is the length of the dispersive medium. This relationship is linearly dependent on the frequency gap from the pump, allowing the photon frequency to be extracted from the time difference between the reference time and the photon's arrival time. By measuring the arrival time of the heralding photon and using Eq. \ref{eq1}, we can determine its frequency and infer the frequency of the corresponding probe photon, which satisfies the energy conservation law.

The probe photons pass through a collimating lens into a free-space diffraction grating, where they are spatially separated by frequency ($\theta \left( f_{\mathrm{pr}} \right)$), as depicted in Fig. \ref{fig1}(d). If targets are present, the photons reflect back and are collected by a telescope, then sent to a detector for arrival time recording.

Fig. \ref{fig1}(e) illustrates how to perform target detection and ranging for objects at different directions and distances. The y-axis target direction ($\theta \left( f_{\mathrm{pr}} \right)$) is derived from the diffraction grating equation and probe photon frequency ($f_{\mathrm{pr}}$), while the target distance $D \left( f_{\mathrm{pr}} \right)$ is obtained from the measured temporal difference ($t_d$) between heralding and probe photons:
\begin{align}
D\left( f_{\mathrm{pr}} \right) &= \frac{c}{2} \left[t_d - 2\left\{ \tau^\prime \left( f_{\mathrm{h}} \right) - \tau^\prime \left( f_{\mathrm{ref}} \right) \right\} \right] \label{eq2}
\end{align}
Here, the raw temporal delay after the dispersive medium and before the diffraction grating is $ 2\left\{ \tau^\prime \left( f_{\mathrm{h}} \right) - \tau^\prime \left( f_{\mathrm{ref}} \right) \right\} = \tau^\prime \left( f_{\mathrm{h}} \right) - \tau^\prime \left( f_{\mathrm{pr}} \right)$. Photon speed in free space is constant at $c$, as dispersion in free space is negligible.

In summary, as the photon-pair generation originates from quantum fluctuations, the probe photon's direction remains unpredictable until the heralding photons' wavelength is measured (See Section 1 of Supplement 1 for frequency randomness). By analyzing the heralding photon's arrival time, we can determine the corresponding probe photon’s wavelength, calculate its direction using the diffraction grating equation, and use the TOF method to find the object’s distance. 

\subsection{Noise resilience theory of QEP-LiDAR}\label{subsec2}

Using the strong correlation between photon pairs, it is well known that quantum LiDAR, quantum imaging, and quantum illumination have greater noise resilience than classical methods. In our QEP-LiDAR method, the noise intensity is defined as the total number of photons measured without probe photons divided by the total number of probe photons measured ($N_{\mathrm{probe : False}} / N_{\mathrm{probe : True}}$), where $N_{\mathrm{Probe : True}}$ and $N_{\mathrm{Probe : False}}$ represent the number of probe photons reflected from the targets and the number of noise photons, respectively. The signal-to-noise ratios (SNR) for the classical (C) and quantum (Q) scenarios are defined as the following equations,
\begin{align}
\mathrm{SNR}_\mathrm{C} &= \frac{N_{\mathrm{SC \left( probe : on | noise : on \right)}} - N_{\mathrm{SC \left( probe : off | noise : on \right)}}}{N_{\mathrm{SC \left( probe : off | noise : on \right)}}}, \label{eq3} \\
\mathrm{SNR}_\mathrm{Q} &= \frac{N_{\mathrm{CC \left( probe : on | noise : on \right)}} - N_{\mathrm{CC \left( probe : off | noise : on \right)}}}{N_{\mathrm{CC \left( probe : off | noise : on \right)}}}, \label{eq4}
\end{align}
where $N_{\mathrm{SC \left( probe : on | noise : on \right)}}$ and $N_{\mathrm{SC \left( probe : off | noise : on \right)}}$ represent the single counts at the probe photon detector when both probe and noise photons are present and when only noise photons are present, respectively. Similarly, $N_{\mathrm{CC \left( probe : on | noise : on \right)}}$ and $N_{\mathrm{CC \left( probe : off | noise : on \right)}}$ denote the coincidence counts between the probe and heralding detectors under the same conditions: with both probe and noise photons present and with only noise photons present, respectively. 

Finally, the SNR enhancement ($E_{\mathrm{SNR}}$), which indicates the degree to which the QEP-LiDAR improves the SNR compared to the classical LiDAR, is defined as
\begin{align}
\mathrm{E}_{\mathrm{SNR}} &= \frac{\mathrm{SNR}_{\mathrm{Q}}}{\mathrm{SNR}_{\mathrm{C}}} = \frac{\nu_{\mathrm{CC}} \eta_{\mathrm{P}} \eta_{\mathrm{H}}}{\left( \nu_{\mathrm{CC}} + \nu_{\mathrm{SC,P}} \right) \eta_{\mathrm{P}} \{ \left( \nu_{\mathrm{CC}} + \nu_{\mathrm{SC,H}} \right)\eta_\mathrm{H} + \nu_{\mathrm{DC,H}} \}} + 1, \label{eq5}
\end{align}
where $\nu_{\mathrm{CC}}$, $\nu_{\mathrm{SC,P}}$, and $\nu_{\mathrm{SC,H}}$ represent the generation rates of photon pairs and noise photons by spontaneous Raman scattering from the photon-pair source at the probe and heralding frequencies~\cite{shin2023photon}, respectively. $\nu_{\mathrm{DC,H}}$ is the dark count rates of the heralding channel detectors. $\eta_{\mathrm{H}}$ and $\eta_{\mathrm{P}}$ are the loss coefficients of the heralding and probe photons, respectively, and $\nu_{\mathrm{noise,P}}$ is the rate of the injected noise photons (See Section 2 of Supplement 1 for the detailed derivation of quantum enhancement).
Interestingly, $\mathrm{E}_{\mathrm{SNR}}$ remains constant regardless of $\nu_{\mathrm{noise,P}}$ and $\eta_{\mathrm{P}}$, suggesting that in noisy and lossy environments where classical systems cannot achieve target detection, the QEP-LiDAR system could serve as a robust solution. Note that this enhancement $\mathrm{E}_{\mathrm{SNR}}$ is similar to the coincidence-to-accidental ratio ($\mathrm{CAR}$) which is defined as follows
\begin{align}
\mathrm{CAR} &= \frac{\nu_{\mathrm{CC}} \eta_{\mathrm{H}}}{\{ \left( \nu_{\mathrm{CC}} + \nu_{\mathrm{SC,P}} \right) + \nu_{\mathrm{DC,P}}/\eta_{\mathrm{P}} \} \{ \left( \nu_{\mathrm{CC}} + \nu_{\mathrm{SC,H}} \right)\eta_\mathrm{H} + \nu_{\mathrm{DC,H}} \}} + 1, \label{eq6}
\end{align}
where $\nu_{\mathrm{DC,P}}$ is the dark count rates of the probe channel detectors. With low dark count detectors, these two values ($\mathrm{E}_{\mathrm{SNR}}$ and $\mathrm{CAR}$) converge, confirming that greater $\mathrm{E}_{\mathrm{SNR}}$ is achieved by effectively exploiting photon pair correlations. This highlights the fundamental role of strong photon correlations in achieving QEP-LiDAR’s superior noise resilience compared to classical systems.

\subsection{Experimental setup}\label{subsec3}

\begin{figure}[htbp]
\centering\includegraphics[width=14cm]{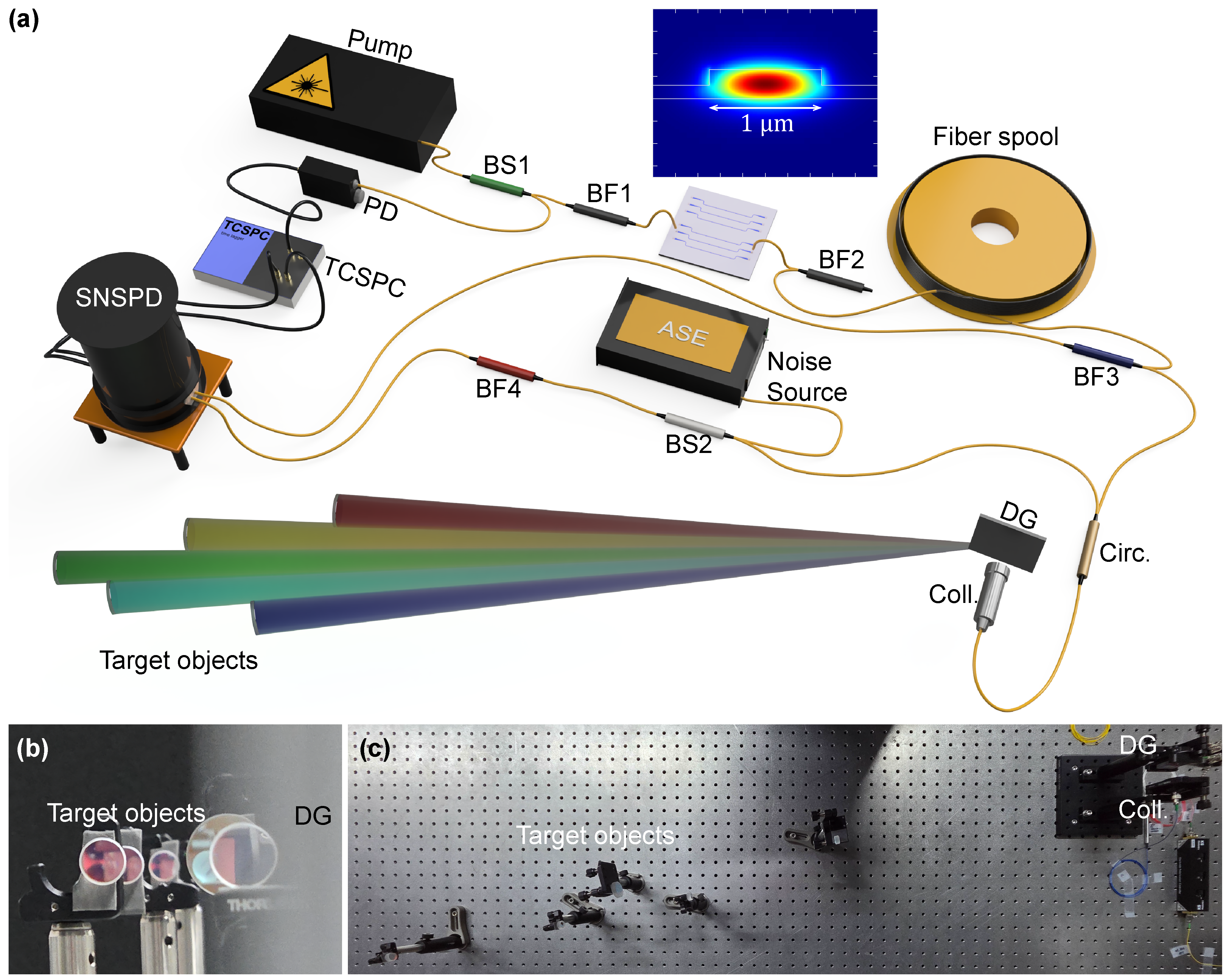}
\caption{Experimental setup. (a) Pump: ps pulsed laser, BS: beam splitter, BF: bandpass filter, Fiber spool: Corning SMF-28 (25.248 km), Noise source: amplified spontaneous emission source, PD: InGaAs photodetector, TCSPC: time-correlated single photon counter, SNSPD: superconducting nanowire single-photon detector, DG: diffraction grating, Circ.: circulator. Coll.: collimating lens. The inset shows the waveguide cross-section and the computed intensity profile of the optical mode. (b) Front view of the targets as seen precisely above the grid. The blurry region on the right side of the image is the DG, which is out of focus. (c) Top view of the targets.}\label{fig2}
\end{figure}

The experimental setup for the QEP-LiDAR is shown in Fig. \ref{fig2}(a). The pump laser operates at a repetition rate of $19.27 \, \mathrm{MHz}$, centered at a frequency of $194.6 \, \mathrm{THz}$ ($1540.56 \, \mathrm{nm}$), with a spectral bandwidth of $31.6 \, \mathrm{GHz}$ ($0.25 \, \mathrm{nm}$) and a pulse duration of $12 \, \mathrm{ps}$. An InGaAs photodetector detects a portion of the pump light,  and a time-correlated single-photon counter (TCSPC, Swabian Instruments) at Ch. 1 equipped with its low jitter option ($\sim$$8 \, \mathrm{ps}$, RMS) records the pulse timing as a reference time ($\tau^\prime_{\mathrm{ref}}$). As a broadband photon-pair source is required for wide-range LiDAR scanning and shorter SpFWM media yield broader photon-pair spectra~\cite{shin2023photon, park2021telecom}, we use a 1-cm silicon waveguide. After passing through the silicon waveguide, the pump light is filtered out by a band-pass filter (BF2), allowing only the photon pairs to enter a dispersive medium (Corning, SMF-28, $25.248 \, \mathrm{km}$ fiber spool), where photons are temporally separated based on their frequencies. The dispersion was measured, resulting in a time separation of approximately $0.4 \, \mathrm{ns/nm}$. Heralding photons are separated from probe photons with a band-pass filter (BF3, central wavelength: $1530 \, \mathrm{nm}$, 0.5 dB bandwidth: $13 \, \mathrm{nm}$) and detected by a superconducting nanowire single-photon detector (SNSPD, Scontel) at Ch. 1. Arrival times are recorded on the TCSPC at Ch. 2. Using the reference and heralding photon arrival times in Eq. \eqref{eq1}, we obtain the frequencies of both the heralding and probe photons via the energy conservation law.

Probe photons are directed into free space through a circulator, collimating lens, and diffraction grating. The collimated photons are spatially dispersed by a diffraction grating (Thorlabs, GR25-0616, $600 \, \mathrm{grooves/mm}$), achieving an angular separation of $0.192 \, \mathrm{deg/nm}$. A portion of photons are reflected by a target, collected by a collimating lens, and routed back through the circulator to an SNSPD at Ch. 2. The photographs of the free-space region in the experimental setup from the diffraction grating to the objects are shown in Figs. \ref{fig2}(b) and \ref{fig2}(c). For signal-to-noise measurements, a broadband noise source (amplified spontaneous emission light source) is combined with the probe photons via a beam splitter (BS2) before detection at the SNSPD. We note that practical receiver collection efficiency limitations for diffusively reflecting targets are common to all LiDAR systems, quantum or classical, rather than a fundamental restriction unique to our method. Future implementations can significantly enhance collection efficiency through optimized optical designs, including larger apertures, multimode fiber coupling, or adaptive optics.

Since the noise photons have a wide wavelength bandwidth, a bandpass filter (BF4, central wavelength: $1551 \, \mathrm{nm}$, 0.5 dB bandwidth: $13 \, \mathrm{nm}$) after the beam splitter (BS2) limits the detected noise to match the spectral range of the probe photons. With noise injected, photons are detected by SNSPD at Ch. 2, and arrival times are recorded on TCSPC at Ch. 3.

\subsection{QEP-LiDAR measurement}\label{subsec4}

\begin{figure}[htbp]
\centering\includegraphics[width=7cm]{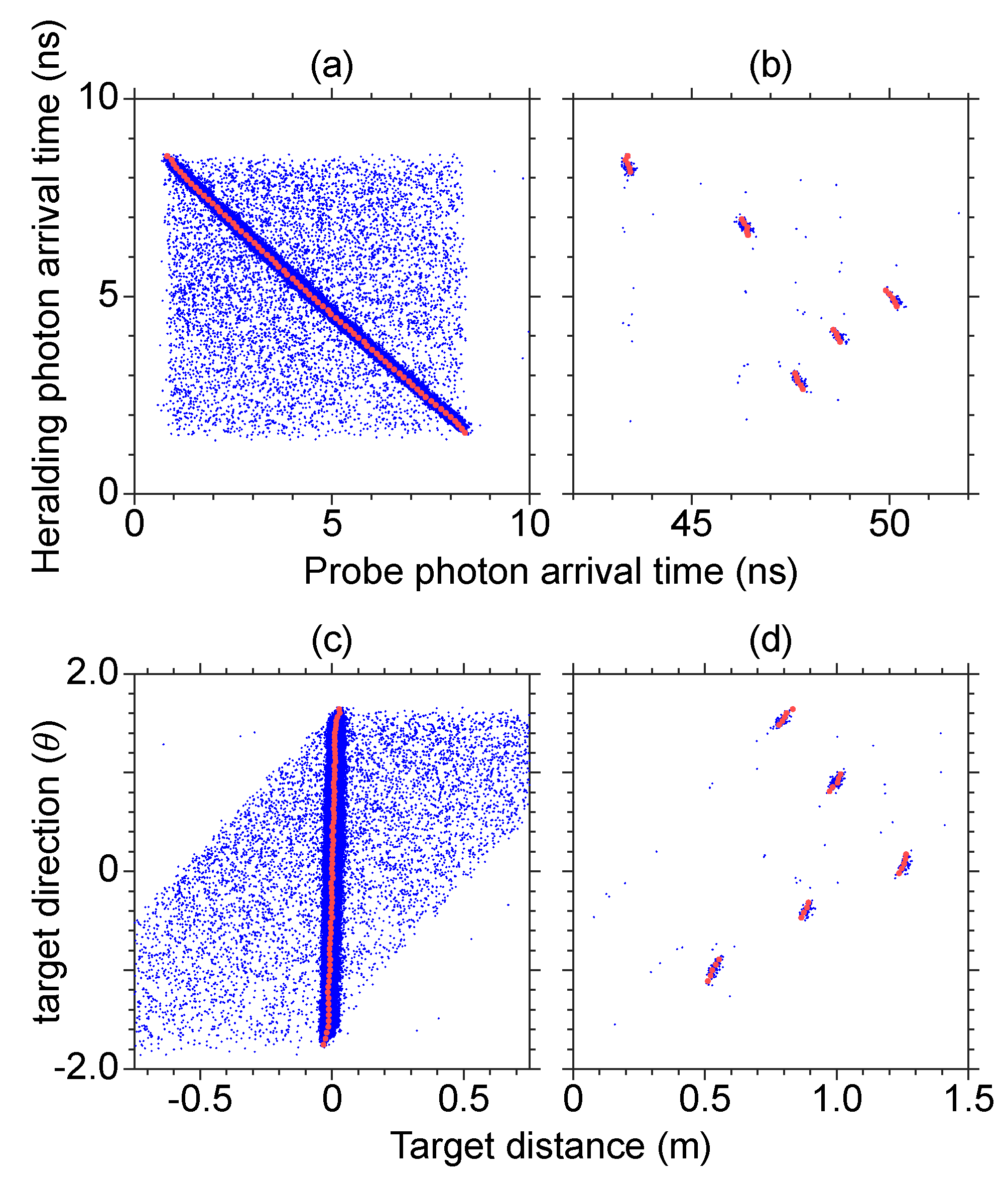}
\caption{Measured coincidence counts of QEP-LiDAR system.  (a)-(b) Scatter plots of CCs with probe photon arrival time on the X-axis and heralding photon arrival time on the Y-axis. (a) when the probe photons are directly detected on SNSPD ch. 2 and, (b) when the probe photons are sent to free space targets. (c)-(d) Scatter plots of CCs with target distance on the X-axis and target direction on the Y-axis. (c) when the probe photons are directly detected on SNSPD ch. 2 and, (d) when the probe photons are sent to free space targets. The maximum CC along the x-axis is marked with a red dot using a Gaussian fit. The data was accumulated for 60 s with a coupled pump peak power of 52.3±1.4 mW.}\label{fig3}
\end{figure}

The distance between the grating and the target is measured using Eq. 2, based on the recorded pump reference and the arrival times of the heralding and probe photons. Coincidence counts (CC) between SNSPD Ch. 1 and 2 are recorded and stored on the TCSPC system. Fig. \ref{fig3}(a) shows the CC results when the output (probe photons) of BF3 in Fig. \ref{fig2} is directly connected to the input of BF4 without free-space transmission or noise injection. The X- and Y-axes represent the arrival times of the probe and heralding photons, respectively. With larger deviation from the pump frequency, heralding photons arrive earlier and probe photons later than the reference time, as shown in Fig. \ref{fig1}(c). This result corresponds to the experimental measurement of the temporal delay between photon pairs and demonstrates a strong frequency-time correlation between heralding and probe photons after propagation through the fiber spool.

The QEP-LiDAR system with five mirror objects was tested using this frequency-time correlation, as shown in Fig. \ref{fig3}(b). Probe photon loss occurs through the collimating lens, grating, and mirrors in the setup shown in Figure \ref{fig2}, resulting in lower overall CC. The arrival time of the probe photon reflected from an object and the distribution of the heralding photon are confined to a specific time range. Similar to Fig. \ref{fig3}(a), the CC distribution is divided by time windows along the y-axis, with the arrival time of the maximum CC value found using a Gaussian fit and marked with a red dot. This enables measurement of the probe photon's flight time, whose wavelength was determined by the heralding photon's arrival time via post-processing.

The QEP-LiDAR results in Fig. \ref{fig3}(b) are presented in terms of the arrival times of heralding and probe photons, but to intuitively understand the spatial distribution of the objects, it is necessary to present them in terms of target distance and direction. For this, we assume that the probe photons are temporally aligned while the heralding photons are distributed in time according to wavelength after the dispersive fiber spool. Thus, the slanted CC distribution in Fig. \ref{fig3}(a) is transformed into the format in Fig. \ref{fig3}(c), assuming simultaneous probe photon arrivals at the grating. This involves shifting the probe arrival time for the maximum CC value to the center and multiplying the speed of light. 
Fig. \ref{fig3}(c), a transformed version of Fig. \ref{fig3}(a), maps the X-axis to the target distance and the Y-axis to the target direction that is estimated using the diffraction grating equation and the probe photon frequency. The X-axis center represents zero target distance. Although the angular observation range demonstrated here is limited to approximately 3.4°, it can be significantly extended by using a diffraction grating with a higher groove density or by increasing the bandwidth of the photon pairs, currently restricted to 13 nm by the CWDM filter.

Similarly, Fig. \ref{fig3}(d) is the transformed format of Fig. \ref{fig3}(b), showing clearly five discrete objects against target distance and direction (See Section 4 of Supplement 1 for Target information and Fig. S3). Target distances are determined by the probe photon arrival peak in each heralding arrival time window. These peak locations match well with the direction and distance of the real targets. Due to spatial constraints on the optical table, the distance between the grating and the sample was set between $0.5\,\,\mathrm{m}$ and $1.5\,\,\mathrm{m}$. The resolution of target distance would be $2.2 \, \, \mathrm{cm}$ by considering a timing jitter of CC (see Section 3 of Supplement 1 for Spatial resolution and Fig. S2), pump pulse bandwidth, and the speed of light in free space, while target direction resolution is $0.144^\circ$, based on timing jitter of heralding photon SC and diffraction grating equation. Although fundamental optical uncertainty principles limit simultaneous angular and axial resolution, in practice, resolution is predominantly constrained by system-specific timing jitter from detectors and electronics, equally affecting quantum and classical approaches.

\subsection{Noise resilience}\label{subsec5}

\begin{figure}[htbp]
\centering\includegraphics[width=7cm]{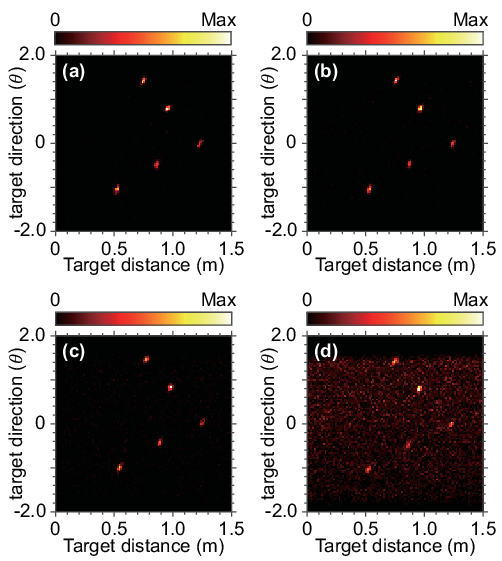}
\caption{Noise resilience experiment. Measured CC distributions for 60 s with target distance on the X-axis and target direction on the Y-axis, (a) when noise is off, (b) when noise is at $+8.2 \, \mathrm{dB}$, (c) when noise is at $+19.0 \, \mathrm{dB}$, and (d) when noise is at $+29.1 \, \mathrm{dB}$.}\label{fig4}
\end{figure}

\begin{figure}[htbp]
\centering\includegraphics[width=14cm]{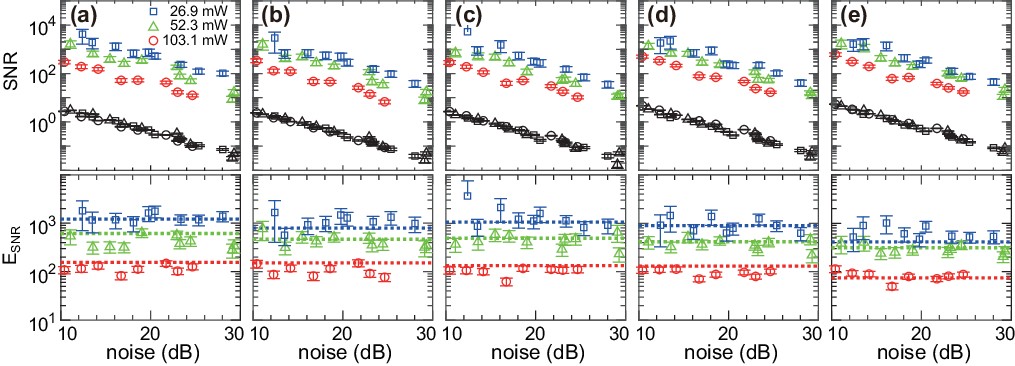}
\caption{$\mathbf{SNR}_\mathbf{Q}$(color), $\mathbf{SNR}_\mathbf{C}$(black), and corresponding $\mathbf{E}_\mathbf{SNR}$ against the noise intensity in various coupled pump peak powers. Signal-to-noise ratio for quantum and classical methods and its quantum enhancement: Blue and black squares: $26.9 \pm 1.4 \, \mathrm{mW}$, green and black triangles: $52.3 \pm 1.4 \, \mathrm{mW}$, and red and black circles: $103.1 \pm 2.9 \, \mathrm{mW}$. (a) target 1, (b) target 2, (c) target 3, (d) target 4 and, (e) target 5. }\label{fig5}
\end{figure}

To demonstrate the noise resilience of the QEP-LiDAR method, we conducted a series of experiments under various noise intensity levels. While the data visualization using blue dots on a white background, as shown in Fig. \ref{fig3}, effectively shows the locations where CCs were measured, it does not indicate the number of CCs detected at each position. To address this, a color-coded map is employed to visually depict the intensity distribution of CCs, as seen in Fig. \ref{fig4}, enabling clear identification of object locations, even against a strong noise background. Fig. \ref{fig4}(a) shows the results in a noise-free scenario, while Fig. \ref{fig4}(b)-(d) present the QEP-LiDAR results measured for 60 s with noise intensity increased to (b)  $8.2 \, \mathrm{dB}$, (c) $19.2 \, \mathrm{dB}$, and (d) $29.1 \, \mathrm{dB}$, compared to the detected probe-photon count rates in Fig. \ref{fig4}(a), still yielding distinct peaks in the true signal CCs. The pump power remained fixed at $52.3 \pm 1.4 \, \mathrm{mW}$. As seen in Fig. \ref{fig4}(d), where noise photons are about a thousand times stronger than the probe photons, noise photons are uniformly distributed along the probe photon arrival time axis. On the y-axis, noise and true signal CCs are distributed within a range of $-1.9^\circ$ to $1.8^\circ$, corresponding to the heralding photon arrival time range from $1.5 \, \mathrm{ns}$ to $8.5 \, \mathrm{ns}$. Therefore, as shown in Fig. \ref{fig4}, target locations remain identifiable even when the noise photons are much stronger than the probe photons.

To quantitatively demonstrate the noise resilience of the QEP-LiDAR method, we compared its SNR to that of a classical method under the same conditions across various noise levels (see Section 6 of Supplement 1 for Quantum enhancement measurement). $\mathrm{SNR}_\mathrm{Q}$ was calculated by selecting the heralding and probe photon arrival time bins with the highest CC count for each target with a $100 \, \mathrm{ps}$ time window, using Eq. \ref{eq4}. $\mathrm{SNR}_\mathrm{C}$ was calculated by applying the same time window used for $\mathrm{SNR}_\mathrm{Q}$ to all the signals from SNSPD Ch.2 over the accumulation time, using Eq. \ref{eq3}. Then, $\mathrm{E}_\mathrm{SNR}$ was calculated from these values using Eq. \ref{eq5}, and the results are shown in Fig. \ref{fig5}. To observe the SNR enhancement based on the photon-pair generation rate, we varied the peak power of the pump coupled to the photon-pair source at $26.9 \, \mathrm{mW}$, $52.3 \, \mathrm{mW}$, and $103.1 \, \mathrm{mW}$. The accumulation time was set to $60 \, \mathrm{s}$ for $52.3 \, \mathrm{mW}$ and $103.1 \, \mathrm{mW}$ and $300 \, \mathrm{s}$ for $26.9 \, \mathrm{mW}$. Fig. \ref{fig5}(a)-(e) show the SNR and $\mathrm{E}_\mathrm{SNR}$  for targets 1 to 5, respectively. In Fig. \ref{fig5}, $\mathrm{SNR}_\mathrm{Q}$ and $\mathrm{SNR}_\mathrm{C}$ decrease almost in parallel as noise intensity increases. $\mathrm{SNR}_\mathrm{C}$ decreases along the same line regardless of the pump peak power, as it is inversely proportional to the noise intensity by definition in Eq. \ref{eq3} and independent of the pump peak power. Meanwhile, $\mathrm{SNR}_\mathrm{Q}$ increases as the pump peak power decreases, benefiting from noise resilience through photon-pair correlation, and is expected to be similar to $\rm{CAR}$ from Eqs. \ref{eq5} and \ref{eq6}. The dotted lines in the second row of Fig. \ref{fig5} represent the CAR for the corresponding pump power (see Section 5 of Supplement 1 for CAR measurement). The measured $\mathrm{E}_\mathrm{SNR}$ values are close to these lines, even when the noise intensity changes by nearly $20 \, \mathrm{dB}$.

\section{CONCLUSION}\label{sec3}

In summary, we successfully demonstrated target detection and ranging using the QEP-LiDAR method, which features inherently unpredictable beam steering. This unpredictability was achieved by exploiting the random frequencies of photon pairs generated via spontaneous four-wave mixing. By resolving the arrival times of heralding photons through a dispersive medium, we determined the frequency and propagation direction of the probe photons. A key advantage of quantum LiDAR, its noise resilience, was validated with a signal-to-noise ratio significantly higher than that of classical LiDAR under identical experimental conditions. The experiment with parallel target detection yielded a resolution of $2.2 \, \, \mathrm{cm}$ and an angular resolution of $0.144^\circ$ under our experimental conditions, which can be readily enhanced by using better electronics and grating, respectively.

In general, quantum LiDAR and imaging inherently provide some degree of stealth observation due to their ability to perform measurements under low-light conditions~\cite{malik2010quantum}. The beam steering method proposed in this study fundamentally prevents the prediction of beam direction until the heralding photon is measured, thereby achieving a higher degree of stealth compared to conventional mechanical or electronic beam steering methods. Furthermore, this study presents the first implementation of a parallel observation system within a quantum LiDAR framework. It is widely recognized that increasing the mode separation in time, space, polarization, or spectrum enhances the resilience to noise~\cite{zhang2020multidimensional, blakey2022quantum}. In our QEP-LiDAR system, we utilized spatially separated probe-photon modes for parallel measurements, resulting in an improved signal-to-noise ratio ($\mathrm{E}_\mathrm{SNR}$), as demonstrated in Fig.\ref{fig5}. Because these metrics rely on greater mode separation compared to raster scanning systems, similar improvements can be anticipated in other mode-separation-based applications. Our experiments successfully reduced the average number of photons per measurement direction while preserving noise resilience. Moreover, parallel observation enables one to capture spatial information across multiple pixels simultaneously, significantly enhancing the measurement speed compared to raster-scanning methods. This advantage may be analogous to the fact that successful pseudo-thermal ghost imaging was achieved with fewer random patterns compared to traditional raster scanning methods~\cite{katz2009compressive}. Further studies are necessary. The combination of inherent unpredictability, high signal-to-noise ratio, and rapid information acquisition demonstrated here holds substantial potential for advancing quantum LiDAR, random number generation, quantum communication, and quantum networking applications.

\begin{backmatter}

\bmsection{Funding}
This research was supported by the Future Challenge Defense Technology Research and Development Program through the Agency for Defense Development (ADD) grant funded by the Defense Acquisition Program Administration (DAPA) in 2023 (No. 912910601) and by the IITP(Institute of Information \& Communications Technology Planning \& Evaluation)-ITRC(Information Technology Research Center) grant funded by the Korea government(Ministry of Science and ICT)(IITP-2025-RS-2022-00164799).

\bmsection{Acknowledgment}
We thank Prof. Wookjae Lee, Prof. Yoon-ho Kim, Jiyoung Kim, and Dr. Sebae Park for their helpful discussions and suggestions.

\bmsection{Disclosures}
The authors declare no conflicts of interest.

\bmsection{Supplemental document}
See Supplement 1 for supporting contents.

\end{backmatter}

%%%%%%%%%%%%%%%%%%%%%%% References %%%%%%%%%%%%%%%%%%%%%%%%%

%%%%%%%%%% If using BibTeX:
\bibliography{references}

\end{document}

\maketitle

\section{Frequency randomness}\label{sec1}
\begin{figure}[ht!]
\centering
\includegraphics[width=0.9\textwidth]{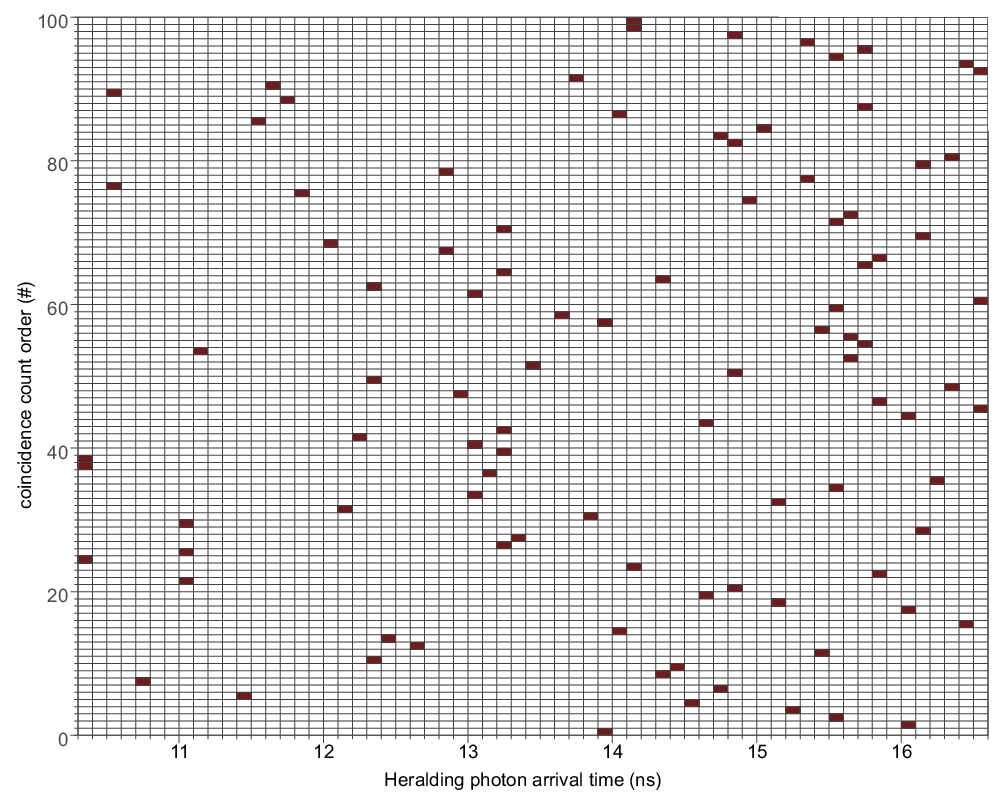}
\caption{\textbf{Inherently unpredictable heralding photon arrival time.} Heralding photon arrival times for each coincidence count, up to the 100th order, in the experimental data used for the coincidence-to-accidental ratio (CAR) measurement.} \label{figS1}
\end{figure}
The two-photon state generated by spontaneous four-wave mixing in a silicon waveguide is expressed as follows: 
\begin{align}
| \Psi \rangle &\propto \iint d\omega_{h}d\omega_{pr}\exp{\left(-4\ln{2}\frac{\left( 2\omega_{p}-\omega_{h}-\omega_{pr} \right)^2}{\left( \delta\omega_{p} \right)^2}+\frac{i \kappa l}{2}\right)}\mathrm{sinc}{\left( \frac{\kappa l}{2} \right)} \hat{a}^{\dagger}_{h}\hat{a}^{\dagger}_{pr}|vac\rangle \label{eqS11}
\end{align}
where $\omega_{p}$, $\omega_{h}$, and $\omega_{pr}$ represent the frequencies of the pump, heralding, and probe photons, respectively. $\Delta\omega_p$ is the spectral bandwidth of pump laser, $\kappa$ is the phase matching condition and $l$ is the length of silicon waveguide. $\hat{a}^{\dagger}_{h}$and $\hat{a}^{\dagger}_{pr}$ are the creation operators of the heralding and probe photons, respectively. Subsequently, the joint spectral intensity ($\mathrm{JSI}$) is given by:
\begin{align}
\mathrm{JSI} = \mathrm{sinc}^2\left(\frac{\kappa l}{2}\right)\exp{\left(-8\ln{2}\frac{\left( 2\omega_{p}-\omega_{h}-\omega_{pr} \right)^2}{\left( \delta\omega_{p} \right)^2}\right)}\label{eqS12}
\end{align}
This represents the probability distribution of photon pair generation in the spectral domain. Furthermore, this demonstrates that the generation of photon pairs follows this probability distribution in an inherently unpredictable and stochastic process. In QEP-LiDAR experiments, this unpredictability in the spectral domain is resolved in the temporal domain with a dispersive medium. Figure \ref{figS1} illustrates this concept, where the heralding photon arrival times are randomly distributed for each coincidence count.

\section{Detailed derivation of quantum enhancement}\label{sec2}

When noise photons are introduced, the detection probabilities for the single count (SC) in the probe channel and the coincidence count (CC) between the probe and heralding channels are expressed by the following relations:
\begin{align}
p_\mathrm{SC \left(\mathrm{probe: on | noise:on}\right)} &= \left( \nu_{\mathrm{CC}} + \nu_{\mathrm{SC,P}} \right)\eta_p+\nu_{\mathrm{noise,P}} + \nu_{\mathrm{DC,P}}, \label{eqS21}\\
p_\mathrm{SC \left(\mathrm{probe: off | noise:on}\right)} &= \nu_{\mathrm{noise,P}} + \nu_{\mathrm{DC,P}}, \label{eqS22}\\
p_\mathrm{CC \left(\mathrm{probe: on | noise:on}\right)} &= \nu_{\mathrm{CC}}\eta_\mathrm{P}\eta_{\mathrm{H}}+\{ \left( \nu_{\mathrm{CC}}+\nu_{\mathrm{SC,P}}\right)\eta_{\mathrm{P}}+\left(\nu_{\mathrm{noise,P}} + \nu_{\mathrm{DC,P}}\right) \}\nonumber\\
&\times\{ \left( \nu_{\mathrm{CC}} + \nu_{\mathrm{SC,H}} \right)\eta_{\mathrm{H}}+\nu_{\mathrm{DC,H}} \},\label{eqS23}\\
p_\mathrm{CC \left(\mathrm{probe: off | noise:on}\right)} &= \left(\nu_{\mathrm{noise,P}} + \nu_{\mathrm{DC,P}}\right)\{ \left( \nu_{\mathrm{CC}} + \nu_{\mathrm{SC,H}} \right)\eta_{\mathrm{H}}+\nu_{\mathrm{DC,H}}\}, \label{eqS24}
\end{align}
where (probe:on$|$noise:on) and (probe:off$|$noise:on) mean the scenarios in which the probe channel is connected or disconnected from the photon pair source. In this notation, x(probe:y$|$noise:z) (where $\mathrm{x}\in\{\mathrm{SC},\mathrm{CC}\}$, $\mathrm{y}\in\{\mathrm{on},\mathrm{off}\}$, $\mathrm{z}\in\{\mathrm{on},\mathrm{off}\}$) denotes the specific measurement configuration. Here, SC and CC stand for single and coincidence count detections, respectively, while on and off refer to whether the probe or noise photons are introduced (on) or blocked (off).
$\nu_{\mathrm{CC}}$ denotes the photon pair generation rate. $\nu_{\mathrm{SC,P}}$ and $\nu_{\mathrm{SC,H}}$ represent the noise photon rates that are generated by the various nonlinear phenomena in the photon pair source at the probe and heralding frequencies, respectively. $\nu_{\mathrm{DC,P}}$ and $\nu_{\mathrm{DC,H}}$ are the dark count rates of the probe and heralding channel detectors. $\eta_{\mathrm{P}}$ and $\eta_{\mathrm{H}}$ are the loss coefficients of the probe and heralding photons, respectively, and $\nu_{\mathrm{noise,P}}$ is the rate of the injected noise photons.

The signal-to-noise ratio (SNR) for the classical (C) and quantum (Q) scenarios are defined by the following equations,
\begin{align}
\mathrm{SNR}_\mathrm{C} &= \frac{N_{\mathrm{SC \left( probe: on | noise: on \right)}} - N_{\mathrm{SC \left( probe: off | noise: on \right)}}}{N_{\mathrm{SC \left( probe: off | noise: on \right)}}}\nonumber\\&= \frac{\left( \nu_{\mathrm{CC}} + \nu_{\mathrm{SC,P}} \right)\eta_{\mathrm{P}}}{\nu_{\mathrm{noise,P}} + \nu_{\mathrm{DC,P}}},\label{eqS25} \\
\mathrm{SNR}_\mathrm{Q} &= \frac{N_{\mathrm{CC \left( probe: on | noise: on \right)}} - N_{\mathrm{CC \left( probe: off | noise: on \right)}}}{N_{\mathrm{CC \left( probe: off | noise: on \right)}}}\nonumber\\&=\frac{\nu_{\mathrm{CC}}\eta_\mathrm{P}\eta_{\mathrm{H}}+\left( \nu_{\mathrm{CC}}+\nu_{\mathrm{SC,P}}\right)\eta_{\mathrm{P}}\{ \left( \nu_{\mathrm{CC}} + \nu_{\mathrm{SC,H}} \right)\eta_{\mathrm{H}}+\nu_{\mathrm{DC,H}} \}}{\left(\nu_{\mathrm{noise,P}} + \nu_{\mathrm{DC,P}}\right)\{ \left( \nu_{\mathrm{CC}} + \nu_{\mathrm{SC,H}} \right)\eta_{\mathrm{H}}+\nu_{\mathrm{DC,H}}\}}, \label{eqS26}
\end{align}
where $N_{\mathrm{SC \left( probe: on | noise: on \right)}}$ and $N_{\mathrm{SC \left( probe: off | noise: on \right)}}$ represent the single counts at the probe photon detector when both the probe and noise photons are present and when only the noise photons are present, respectively. Similarly, $N_{\mathrm{CC \left( probe: on | noise: on \right)}}$ and $N_{\mathrm{CC \left( probe: off | noise: on \right)}}$ denote the coincidence counts between the probe and heralding detectors under the same conditions: when both the probe and noise photons present and when only noise photon present, respectively. 

 Finally, the SNR enhancement ($\mathrm{E}_{\mathrm{SNR}}$) of the QEP-LiDAR compared to classical LiDAR is given by:
\begin{align}
\mathrm{E}_{\mathrm{SNR}} &= \frac{\mathrm{SNR}_{\mathrm{Q}}}{\mathrm{SNR}_{\mathrm{C}}}\nonumber\\ &= \frac{\nu_{\mathrm{CC}} \eta_{\mathrm{P}}\eta_{\mathrm{H}}+\left( \nu_{\mathrm{CC}} + \nu_{\mathrm{SC,P}} \right)\eta_{\mathrm{P}} \{ \left( \nu_{\mathrm{CC}} + \nu_{\mathrm{SC,H}} \right)\eta_\mathrm{H} + \nu_{\mathrm{DC,H}} \}}{\left( \nu_{\mathrm{CC}} + \nu_{\mathrm{SC,P}} \right)\eta_{\mathrm{P}} \{ \left( \nu_{\mathrm{CC}} + \nu_{\mathrm{SC,H}} \right)\eta_\mathrm{H} + \nu_{\mathrm{DC,H}} \}}\nonumber\\ &= \frac{\nu_{\mathrm{CC}} \eta_{\mathrm{H}}}{\left( \nu_{\mathrm{CC}} + \nu_{\mathrm{SC,P}} \right) \{ \left( \nu_{\mathrm{CC}} + \nu_{\mathrm{SC,H}} \right)\eta_\mathrm{H} + \nu_{\mathrm{DC,H}} \}} + 1. \label{eqS27}
\end{align}

\section{Spatial resolution}\label{sec3}
\begin{figure}[ht!]
\centering
\includegraphics[width=0.9\textwidth]{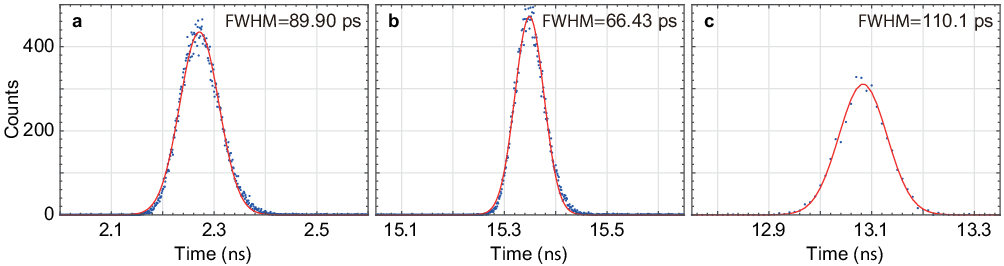}
\caption{\textbf{Timing jitter measurement.} \textbf{a,} Single counts histogram of the heralding channel. \textbf{b,} Single counts histogram of the probe channel. \textbf{c,} Coincidence counts histogram between the heralding and probe channels} \label{figS3}
\end{figure}
For the derivation of spatial resolution, the timing jitter (full width at half maximum, FWHM) of single and coincidence counts in the heralding and probe channels was measured simultaneously using photon pairs generated from a silicon waveguide. Fig. \ref{figS3} presents the following results: The timing jitter of single counts in the heralding channel ($\delta\tau_{\mathrm{SC},h}$) is $89.90\,\mathrm{ps}$, that in the probe channel ($\delta\tau_{\mathrm{SC},pr}$) is $66.43\,\mathrm{ps}$, and the timing jitter of coincidence counts between them ($\delta\tau_{\mathrm{CC}}$) is $110.1\,\mathrm{ps}$.

The target distance resolution ($\delta D$) is calculated as follows:
\begin{align}
\delta D = \frac{c}{2}\sqrt{\left(\delta\tau_{\mathrm{CC}}\right)^2+\left(\mathrm{D}_\lambda L\, \delta\omega_{p} \right)^2}=2.2\,\mathrm{cm},\label{eqS31}
\end{align}
where $D_{\lambda}$ is the dispersion parameter of the fiber spool, $L$ is the length of fiber spool and $\delta\omega_{p}$ is the pump bandwidth. The resulting target-distance resolution $\delta\mathrm{D}$ is $2.2\,\mathrm{cm}$ using our experimantal parameters, $D_{\lambda}L\approx0.4\,\mathrm{ns/nm}$ and $\delta\omega_{p}=0.25\,\mathrm{nm}$.
%Based on the experimental setup ($D_{\lambda}L\approx0.4\,\mathrm{ns/nm}$ and $\delta\omega_{p}=0.25\,\mathrm{nm}$), $\delta D$ is calculated to be $2.2\, \mathrm{cm}$.

The diffraction grating equation is given as follows:
\begin{align}
\alpha\left(\sin{\theta_{m}}+\sin{\theta_{\mathrm{i}}}\right)=m\lambda_{pr},\label{eqS32}
\end{align}
and the corresponding angular separation for the $-1$ diffraction grating order ($m$) is expressed as follows:
\begin{align}
\frac{d\theta_{-1}}{d\lambda_{pr}} =-\frac{1}{\sqrt{\alpha^2-\left(\lambda_{pr}+\alpha\cdot\sin{\theta_{\mathrm{i}}}\right)^2}},\label{eqS33}
\end{align}
In the experimental setup with a reflective diffraction grating (Thorlabs, GR25-0616, $600\,\mathrm{grooves/mm}$), the probe photon angular separation was $0.192 \, \mathrm{deg/nm}$, where the incident angle ($\theta_{\mathrm{i}}$) was $3.05^{\circ}$, the probe channel center wavelength ($\lambda_{pr,0}$) was $1551 \,\mathrm{nm}$ and the diffraction grating period ($\alpha$) was $1.67\,\mu m$.

The target direction resolution ($\delta\theta_{pr}$) is calculated as follows:
\begin{align}
\delta\theta_{pr}&\approx\delta\lambda\times\Big|\frac{d\theta_{-1}}{d\lambda_{pr}}\Big|_{\lambda_{pr,0}=1551 \,\mathrm{nm}}\nonumber\\&\approx\sqrt{\left(\frac{\delta\tau_{\mathrm{SC},h}}{D_{\lambda}L}\right)^2+\left(\frac{\lambda_{pr,0}}{|\mathrm{R}|}\right)^2}\times\Big|\frac{d\theta_{-1}}{d\lambda_{pr}}\Big|_{\lambda_{pr,0}=1551 \,\mathrm{nm}}, \label{eqS34}
\end{align}
where $\mathrm{R}$ is the diffraction grating resolving power, which represents the FWHM value of the point spread function and is calculated as follows:
\begin{align}
|\mathrm{R}|=\frac{\lambda_{pr,0}}{\delta\lambda_{grating}}=\Big|\frac{\alpha}{w}\Big|, \label{eqS34}
\end{align}
where $w$ is the beam waist at the diffraction grating surface. Finally, based on the experimental setup with $w=3.6 \,\mathrm{mm}$, the resulting target direction resolution $\delta\theta_{pr}$ is $0.144^\circ$.

\section{Target information}\label{sec4}
\begin{figure}[ht!]
\centering
\includegraphics[width=0.9\textwidth]{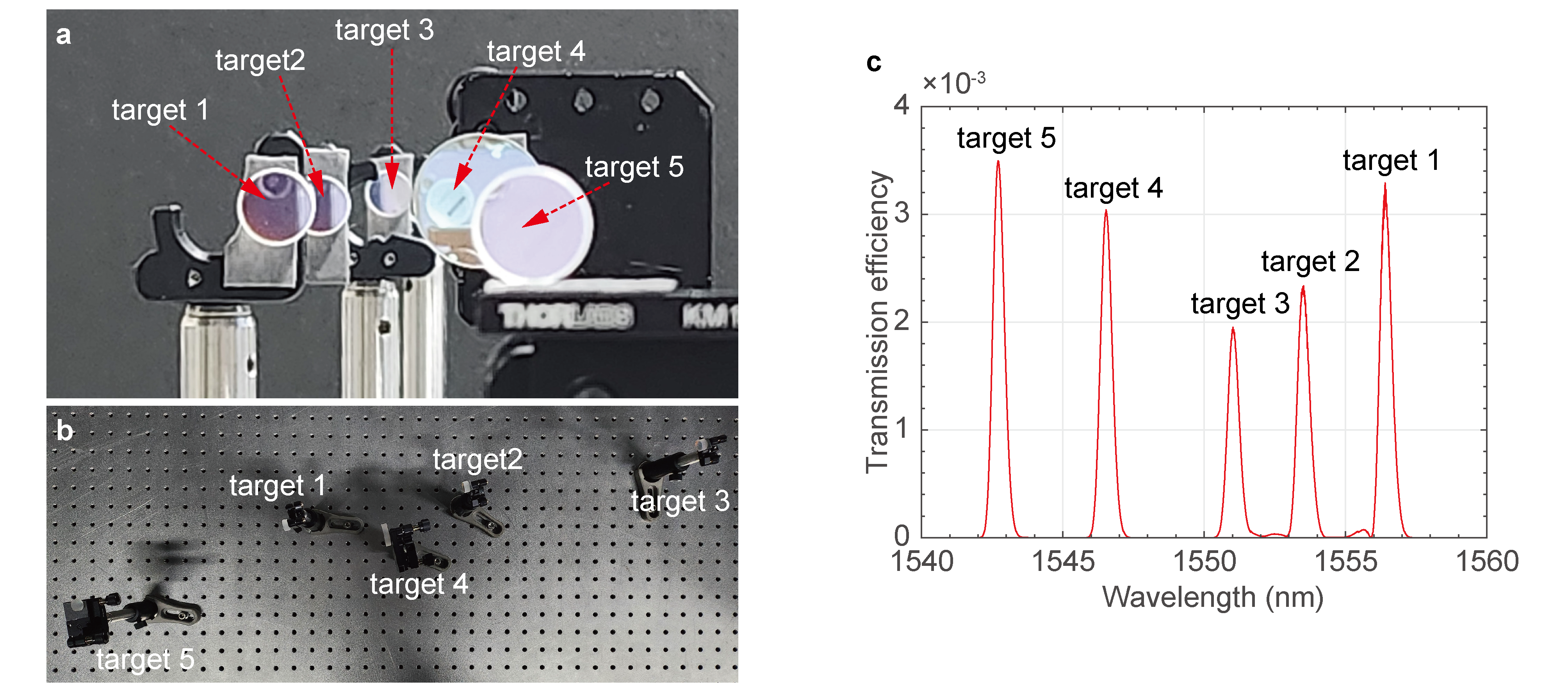}
\caption{\textbf{Target information.} \textbf{a,} Front view of the targets. \textbf{b,} Top view of the targets. \textbf{c,} Wavelength sweep measurement for the five targets.} \label{figS4}
\end{figure}
To calibrate the target positions with the measurement results in the temporal domain, we adjusted the experimental setup in Fig. 2 and performed a spectrum measurement. Specifically, a C-band tunable laser was connected to the input port of the grating coupler, while the output port of the beam splitter was coupled to a power meter. The targets are shown in Fig. \ref{figS4}, with the top view shown in panel (a) and the front view in panel (b). The transmission efficiency spectrum of part of the system is presented in Fig. \ref{figS4}(c), including contributions from the silicon waveguide, bandpass filters, fiber spool, beam splitter, circulator, and collimator. The five peaks in the spectrum correspond to the five targets. Please see the section on 'Calibration' for detailed information on the calibration process.

\section{CAR measurement}\label{sec5}

\begin{figure}[ht!]
\centering
\includegraphics[width=0.9\textwidth]{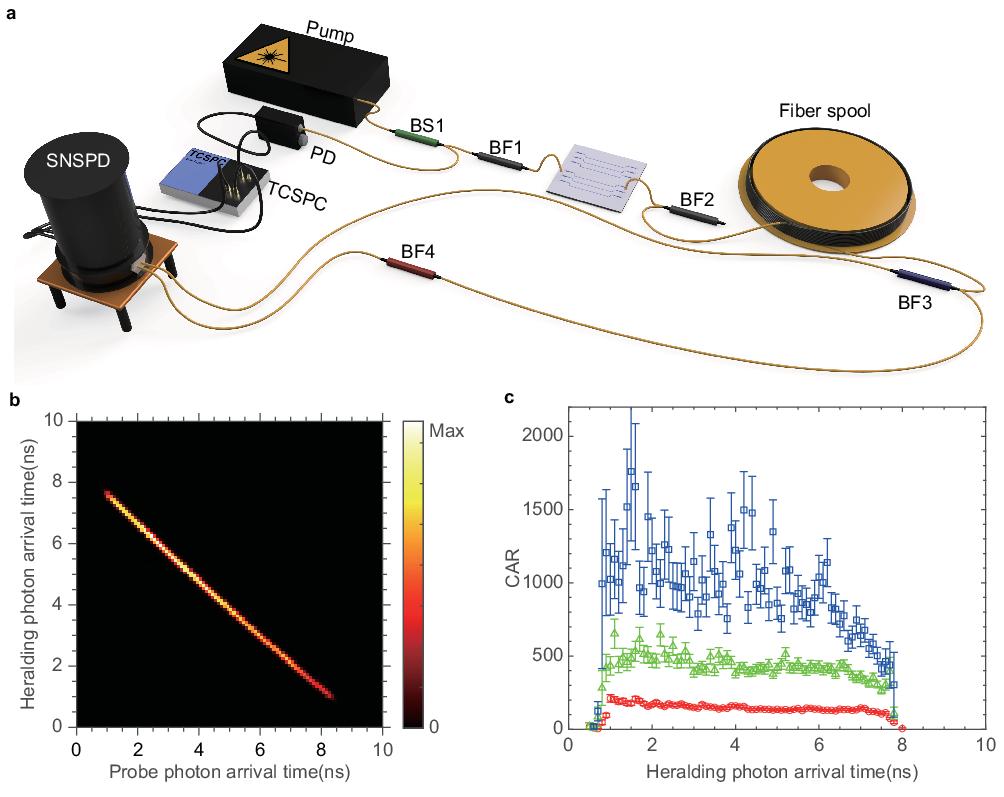}
\caption{\textbf{CAR measurement.} \textbf{a,} Experimental setup for the CAR measurement. \textbf{b,} Joint temporal intensity distribution after photon pairs are temporally separated according to their frequencies, measured with a coupled peak pump power of $51.79 \pm 0.25 \, \mathrm{mW}$ and an accumulation time of $240 \, \mathrm{s}$. Each pixel has a width and height of $100 \, \mathrm{ps}$. \textbf{c,} Measured CAR values for each heralding photon arrival time window ($100\, \mathrm{ps}$) and the corresponding probe photon arrival time window ($100\, \mathrm{ps}$) for various coupled peak pump powers and accumulation times. Blue squares ($26.9 \pm 0.16 \, \mathrm{mW}$, $960\, \mathrm{s}$), green triangles ($51.79 \pm 0.25 \, \mathrm{mW}$, $240\, \mathrm{s}$), and red circles ($100.4 \pm 0.5 \, \mathrm{mW}$, $60\, \mathrm{s}$).} \label{figS5}
\end{figure}

The experimental setup for measuring the coincidence-to-accidental counts ratio (CAR) is shown in Fig. \ref{figS5}(a). Photon pairs are generated via spontaneous four-wave mixing (SpFWM) in a silicon waveguide fabricated on a silicon-on-insulator wafer. The pump laser (Calmer Laser, repetition rate: 19.27 MHz) operates at a center frequency of $194.6 \, \mathrm{THz}$ ($1540.56 \, \mathrm{nm}$) with a spectral bandwidth of $31.6 \, \mathrm{GHz}$ ($0.25 \, \mathrm{nm}$) and a corresponding pulse duration of $12 \, \mathrm{ps}$. The silicon waveguide is $1.0 \, \mathrm{cm}$ long and designed to optimize both broadband spectral bandwidth and high generation rate. Before entering the silicon waveguide, the pump laser is split into a reference signal using a beam splitter (BS1), detected by an InGaAs photodetector (PD, Thorlabs), and recorded on a time-correlated single-photon counter (TCSPC, Swabian Instruments) Ch. 1 as a reference for photon pair arrival times.

After passing through the silicon waveguide, the pump laser is filtered out by a bandpass filter (BF2), leaving only the generated photon pairs to be directed into a fiber spool (Corning, SMF-28, $25.248 \, \mathrm{km}$), which serves as the dispersive medium for the temporal separation of photon pairs. The dispersion caused by the fiber spool is measured based on the photon wavelength and its corresponding arrival time, resulting in approximately $0.4 \, \mathrm{ns/nm}$. The heralding photons are separated from the probe photons using a bandpass filter (BF3, CWDM, central wavelength: $1530 \, \mathrm{nm}$, bandwidth: $13\, \mathrm{nm}$) and sent to the superconducting nanowire single-photon detector (SNSPD, Scontel) Ch.1, with their arrival times recorded on TCSPC Ch.2. The probe photons are directed to SNSPD Ch.2 after passing through a bandpass filter (BF4, CWDM, central wavelength: $1551\, \mathrm{nm}$, bandwidth: $13\, \mathrm{nm}$), and their arrival times are recorded on TCSPC Ch.2.

Figure \ref{figS5}(b) shows the joint temporal intensity after the fiber spool, where the photon pairs are temporally separated according to their frequencies. The measurement was conducted with a coupled peak pump power of $51.79 \,\mathrm{mW}$ and an accumulation time of $240 \,\mathrm{s}$. The coincidence counts are stretched along an apparent diagonal line, indicating a temporally resolved spectral correlation between the photon pairs. The CAR is defined as follows:
\begin{align}
\mathrm{CAR}&=\frac{N_{\mathrm{CC}}}{N_{\mathrm{ACC}}}\nonumber\\
&= \frac{\nu_{\mathrm{CC}} \eta_{\mathrm{P}}\eta_{\mathrm{H}}+\{\left( \nu_{\mathrm{CC}} + \nu_{\mathrm{SC,P}} \right)\eta_{\mathrm{P}}+\nu_{\mathrm{DC,P}}\} \{ \left( \nu_{\mathrm{CC}} + \nu_{\mathrm{SC,H}} \right)\eta_\mathrm{H} + \nu_{\mathrm{DC,H}} \}}{\{\left( \nu_{\mathrm{CC}} + \nu_{\mathrm{SC,P}} \right)\eta_{\mathrm{P}}+\nu_{\mathrm{DC,P}}\} \{ \left( \nu_{\mathrm{CC}} + \nu_{\mathrm{SC,H}} \right)\eta_\mathrm{H} + \nu_{\mathrm{DC,H}} \}}\label{eqS51}
\end{align}
where $N_{\mathrm{CC}}$ is the number of coincidence counts, $N_{\mathrm{ACC}}$ is the number of accidental counts, and $\nu_{\mathrm{DC,P}}$ is the dark count rates of the probe channel detectors. $N_{\mathrm{CC}}$ was determined by segmenting the heralding photon arrival time (y-axis) into $100 \,\mathrm{ps}$ intervals, identifying the peak position of the probe photon arrival time (x-axis) through Gaussian fitting of each coincidence count distribution, and counting the coincidence events within a 100 ps window centered at the peak position in both temporal dimensions. $N_{\mathrm{ACC}}$ was calculated by averaging the counts obtained from multiple time windows of the same size as those used for coincidence counting, with these windows shifted by intervals corresponding to the pump pulse period from the peak position.

Figure \ref{figS5}(c) is the CAR for each heralding photon arrival time segment. For each coupled peak pump power of $100.4 \,\mathrm{mW}$, $51.19 \,\mathrm{mW}$, $26.52 \,\mathrm{mW}$ accumulation times are set as $60 \,\mathrm{s}$, $240 \,\mathrm{s}$, $960 \,\mathrm{s}$ in order to match total coincidence counts considering that the photon pair generation rate in SpFWM is proportional to the square of the peak pump power. As the frequency of the generated photon pairs moves further away from the pump photon frequency, i.e., as the heralding photon arrival time decreases, the phase-matching condition of SpFWM deteriorates, resulting in a decrease in $\nu_{CC}$ and a corresponding increase in CAR. This trend is well confirmed in Fig. \ref{figS5}(c).

\section{Quantum enhancement measurement}\label{sec6}

\begin{figure}
\centering
\includegraphics[width=0.9\textwidth]{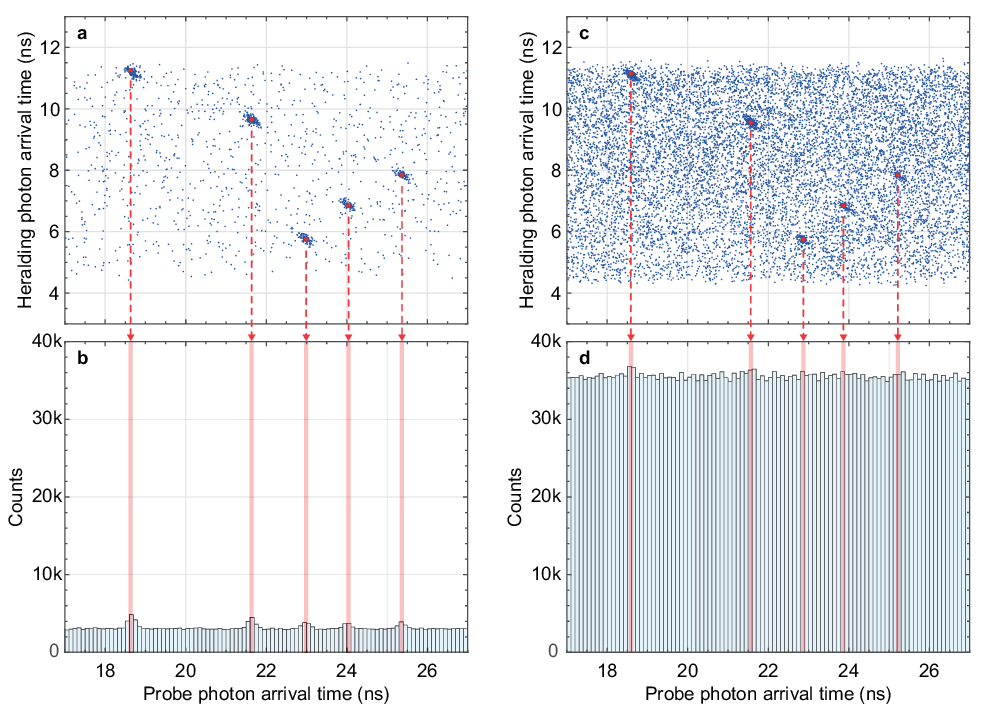}
\caption{\textbf{Quantum enhancement measurement.} \textbf{(a,c)} QEP-LiDAR ${\mathrm{CC \left( probe: on | noise: on \right)}}$ raw data (blue dots) and target detection results (red dots), \textbf{(b,d)} classical LiDAR ${\mathrm{SC \left( probe: on | noise: on \right)}}$ raw data histogram (blue, bin width: $100\,\mathrm{ps}$). Measurements were performed at a coupled peak pump power of ${52.31\pm 1.35  \, \mathrm{mW}}$ under various noise intensities: \textbf{a-b}: $+20.45\,\mathrm{dB}$ and \textbf{c-d}: $+34.17\,\mathrm{dB}$.}\label{figS6}
\end{figure}

For a fair comparison between $\mathrm{SNR}_{\mathrm{C}}$ and $\mathrm{SNR}_{\mathrm{Q}}$, measurements were conducted simultaneously with the same probe photons and targets. Without information about the heralding photon, there is absolutely no way to determine the diffraction direction of the probe photon diffracted by the diffraction grating, inducing the inherent unpredictability. In QEP-LiDAR measurements, the heralding photon - the quantum counterpart of the probe photon - provides this critical information: the wavelengths of the heralding photon and its corresponding probe photon due to the energy conservation law. Using the dispersion medium's known dispersion properties, the probe photon's arrival time with no free-space flight can be identified, and finally, the distance to the object can be calculated using the difference between this arrival time and the actual measured time. 

In classical LiDAR, target detection is proceeded by the temporal delay measurement between the reference signal and probe photon (which is the same as conventional ToF-LiDAR). Additionally, measuring the distance to an object requires the assumption that the wavelength of the probe photon is known. Therefore, it can be assumed that the probe photon of classical LiDAR $N_{\mathrm{probe: True}}$ reaches the QEP-LiDAR's probe photon arrival time in the same experimental environment, while erroneous probe photons $N_{\mathrm{probe: false}}$ are assumed to arrive uniformly. To ensure distinguishability in the classical LiDAR measurements, the distances to each target were adjusted to prevent overlap between probe photon arrival times from different targets.

Figure \ref{figS6}(a) shows the CCs raw data in QEP-LiDAR measurement when the pump power is ${52.31\pm 1.35  \, \mathrm{mW}}$ and the noise intensity is $+20.45\,\mathrm{dB}$. The five red dots represent the highest CC positions for each target. The quantity $N_{\mathrm{CC \left( probe: on | noise: on \right)}}$ for each target is determined by counting the number of CCs within a $100\,\mathrm{ps}$ time window centered on the red dot. Fig. \ref{figS6}(b) presents the histogram of SCs in classical LiDAR measurement conducted simultaneously with the QEP-LiDAR measurements in Fig. \ref{figS6}(a). Here, $N_{\mathrm{SC \left( probe: on | noise: on \right)}}$ for each target is calculated by counting the SCs within a $100\,\mathrm{ps}$ time window centered on the x-axis position of the corresponding red dot. The values of $N_{\mathrm{SC \left( probe: off | noise: on \right)}}$ and $N_{\mathrm{CC \left( probe: off | noise: on \right)}}$ are derived using the same time window for the $\left( \mathrm{probe: off | noise: on} \right)$ case for each target.

Figure \ref{figS6}(c-d) display the raw CC data from QEP-LiDAR and the raw SC data from classical LiDAR when the pump power is ${52.31\pm 1.35 \, \mathrm{mW}}$ and the noise intensity is $+34.17\,\mathrm{dB}$ - approximately 20 times higher than the noise level shown in Fig. \ref{figS6}(a-b). These figures illustrate that classical LiDAR is unable to track target locations as noise photons vastly outnumber probe photons. In contrast, QEP-LiDAR effectively tracks target locations by leveraging the quantum correlation between heralding and probe photons to filter out noise photons.

\section{Calibration}\label{sec7}

\begin{figure}[ht!]
\centering
\includegraphics[width=0.9\textwidth]{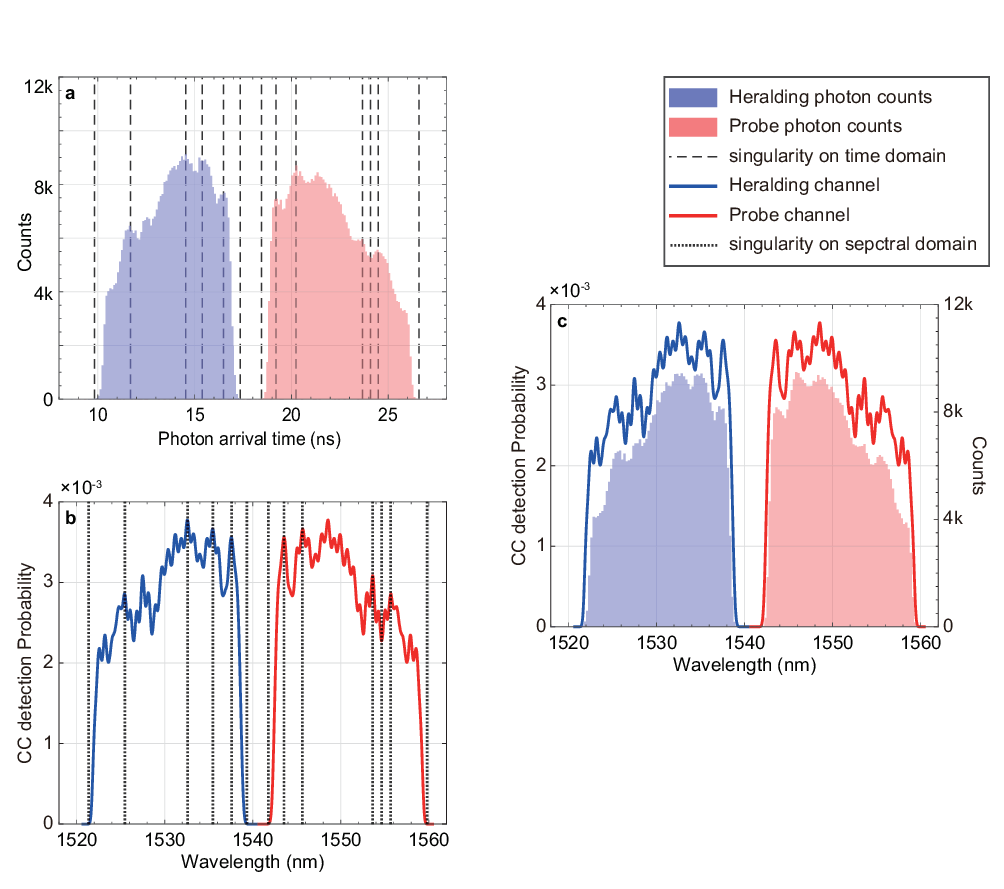}
\caption{\textbf{Calibration process.} \textbf{a}, Histogram of raw coincidence count (CC) data (red area: probe photons, blue area: heralding photons), including singularities (dotted lines) used for calibration. \textbf{b}, Detection probability of CCs (red line: probe channel, blue line: heralding channel), calculated by considering the transmission efficiencies of the probe and heralding channels, with singularities (dotted lines). \textbf{c}, calibration results, illustrating the mapping from photon arrival time to photon wavelength.}\label{figS7}
\end{figure}

Due to the broad spectral range of QEP-LiDAR operation, treating the fiber spool dispersion as a constant is insufficient for accurately converting photon arrival time to photon wavelength. To address this, we performed a calibration to map photon arrival times to wavelengths, as shown in Fig. \ref{figS7}. Fig. \ref{figS7}(a) shows the arrival time histogram of probe photons (red) and heralding photons (blue) obtained from CCs measured during the CAR measurement. The uneven distribution in the histogram is attributed to the non-uniform spectral transmittance of the silicon waveguide used for photon pair generation. Fig. \ref{figS7}(b) shows the probability of a generated photon pair being detected as a coincidence count at the SNSPDs. This probability was determined by multiplying $\eta_{\mathrm{P}}$ and $\eta_{\mathrm{H}}$, which were obtained through a wavelength sweep measurement. Singularities, indicated by dashed lines in Fig. \ref{figS7}(a) and Fig. \ref{figS7}(b), exhibit similar characteristics, and calibration was performed by fitting these features to align them during the conversion of photon arrival time to wavelength. Fig. \ref{figS7}(c) presents the calibration result, demonstrating that the photon arrival times, once converted to wavelengths, align accurately with the expected target wavelengths. This calibration ensures precise time-to-wavelength mapping, which is essential for reliable QEP-LiDAR measurements.

\section{Fisher information comparison}\label{sec8}

\begin{figure}[ht!]
\centering
\includegraphics[width=0.9\textwidth]{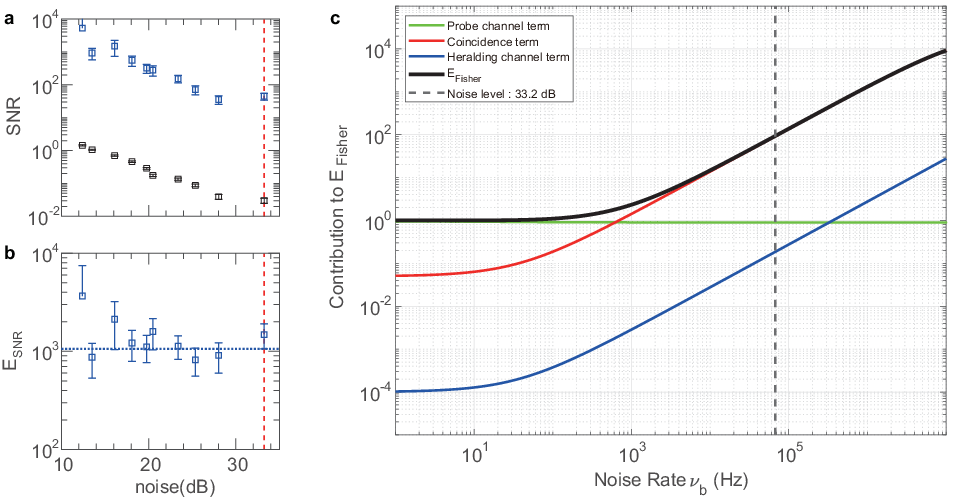}
\caption{\textbf{Fisher information enhancement on target 3 measurements} \textbf{a}, Measured  $\mathrm{SNR_Q}$ (black squares) and $\mathrm{SNR_Q}$ (blue squares) where the pump power 26.9 ± 1.4mW, \textbf{b}, $\mathrm{E_{SNR}}$ (blue square) and corresponding CAR (dotted blue line), \textbf{c}, $\mathrm{E_{Fisher}}$ (black line) of the probe's transmission toward target 3 and individual terms in $\mathrm{E_{Fisher}}$ (colored lines).}\label{figS8}
\end{figure}

QEP-LiDAR employs phase-insensitive coincidence counting measurements. In this regime, the ranging accuracy is dominated by the timing jitter of the photon detection system rather than the fundamental correlation time between photon pairs. As a result, the quantum advantages typically observed over uncertainty-limited classical Gaussian states do not directly apply here. Nonetheless, it has a clear advantage in probe transmission estimation in noisy environments over its practical counterpart, the probe intensity equivalent classical photon counting measurement system.

Fisher information quantifies the amount of information about an unknown parameter that can be obtained from measured data as a numerical value, and it is derived through the maximal likelihood estimation of the joint probability distribution for the corresponding measurement\cite{liu2019enhancing}.

The measurement data for each spatial mode of QEP-LiDAR consists of the photon counts measured by single-photon detectors in the heralding channel and probe channel, $N_{\mathrm{P}}$ and $N_{\mathrm{H}}$, and the coincidence count between the two detectors, $N_{\mathrm{CC}}$. When noise $\nu_b$ per second is injected into the probe channel, the rate at which $N_{\mathrm{CC}}$, $N_{\mathrm{H}}$, and $N_{\mathrm{P}}$ are measured for each pump pulse is as follows
\begin{align}
&\mathrm{P_{CC}} = \frac{\nu_\mathrm{CC} \eta_\mathrm{H} \eta_\mathrm{P} + \nu_\mathrm{CC} \eta_\mathrm{H} \nu_b T_\mathrm{CC}}{T_{pump}}= \nu\eta_\mathrm{H}\eta_\mathrm{P} + \nu\eta_\mathrm{H}\nu_b T_\mathrm{CC}, \label{eqS81}\\
&\mathrm{P_{P}} = \frac{\nu_\mathrm{CC} \eta_\mathrm{P} + \nu_b T_{pump}}{T_{pump}} - \mathrm{P_{CC}} = \nu\eta_P + \nu_b  - \mathrm{P_{CC}},  \label{eqS82}\\
&\mathrm{P_{H}}=\frac{\nu_\mathrm{CC} \eta_\mathrm{H}}{T_{pump}} - \mathrm{P_{CC}} = \nu\eta_H - \mathrm{P_{CC}},  \label{eqS83}
\end{align}
where $\nu$ is the photon pair generation rate per second and $T_{pump}$ is the pump laser period, respectively. The joint probability distribution $P(N_{\mathrm{CC}}, N_\mathrm{H}, N_\mathrm{P}; T_{pump})$ is as follows
\begin{align}
P(N_{\mathrm{CC}}, N_\mathrm{H}, N_\mathrm{P}) = f(N_{\mathrm{CC}}, \mathrm{P_{CC}} T_{pump})f(N_{\mathrm{H}}, \mathrm{P_{H}} T_{pump})f(N_{\mathrm{P}}, \mathrm{P_{P}} T_{pump}),   \label{eqS84}
\end{align}
where each count follows a Poisson distribution $f(N,\lambda) = \frac{\lambda^N \exp(-\lambda)}{N!}$, respectively. The Fisher information for the probe transmission parameter $\eta_\mathrm{P}$ is as follows
\begin{align}
I_Q &= \sum_{N_\mathrm{CC},N_\mathrm{H},N_\mathrm{P}}^{+\infty} P(N_\mathrm{CC}, N_\mathrm{H}, N_\mathrm{P}; T_{\text{pump}}) \left(\frac{\partial}{\partial\eta_\mathrm{P}} \log P(N_\mathrm{CC}, N_\mathrm{H}, N_\mathrm{P}; T_{\text{pump}})\right)^2\nonumber\\
&= \nu^2\left(\frac{\eta_\mathrm{H}^2}{\mathrm{P_{CC}}} + \frac{\eta_\mathrm{H}^2}{\mathrm{P_H}} + \frac{(1-\eta_\mathrm{H})^2}{\mathrm{P_P}}\right) T_{\text{pump}}.\label{eqS85}
\end{align}

For the classical counterpart operating in the same spatial mode (with $\eta_\mathrm{H}=0$), the probability distribution and Fisher information are as follows
\begin{align}
P(N_\mathrm{P}; T_{\text{pump}})|_{\eta_H=0} = f(N_{\mathrm{P}}, \mathrm{P_{P}} T_{pump})|_{\eta_\mathrm{H}=0},\label{eqS86}\\
I_C = T_{\text{pump}} \left(\frac{\nu^2}{\mathrm{P_P}}\right)\bigg|_{\eta_\mathrm{H}=0} = \frac{\nu^2}{\nu\eta_\mathrm{P} + \nu_b} T_{\text{pump}}.\label{eqS87}
\end{align}
Consequently, the Fisher information enhancement $\mathrm{E_{Fisher}}$ is as follows
\begin{align}
\mathrm{E_{Fisher}}= \frac{I_Q}{I_C} =\left(\frac{\eta_\mathrm{H}^2}{\mathrm{P_{CC}}} + \frac{\eta_\mathrm{H}^2}{\mathrm{P_H}} + \frac{(1-\eta_\mathrm{H})^2}{\mathrm{P_P}}\right) (\nu\eta_\mathrm{P} + \nu_b).\label{eq88}
\end{align}

Fig. \ref{figS8}(a,b) presents $\mathrm{SNR_Q}$, $\mathrm{SNR_C}$, and $\mathrm{E_{SNR}}$ measured across various noise intensities at a pump power of 26.9 ± 1.4 mW. Fig. \ref{figS8}(c) illustrates $\mathrm{E_{Fisher}}$ and the individual contributions of each term in Eq. \ref{eq88} as a function of $\nu_b$ under identical experimental conditions.

As demonstrated in Fig. \ref{figS8}(c), the first term (coincidence term) increasingly dominates $\mathrm{E_{Fisher}}$ with rising noise intensity. This occurs because the strong photon-pair correlations significantly suppress the $\nu_b$ contribution to this term, unlike the other terms. At the maximum experimental noise intensity of 33.2 dB, $\mathrm{E_{Fisher}}$ reaches approximately 100, indicating that coincidence counting enables target detection 100 times faster in such high-noise environments.

% Bibliography
\bibliography{references}

%Manual citation list
%\begin{thebibliography}{1}
%\bibitem{Zhang:14}
%Y.~Zhang, S.~Qiao, L.~Sun, Q.~W. Shi, W.~Huang, %L.~Li, and Z.~Yang,
 % \enquote{Photoinduced active terahertz metamaterials with nanostructured
  %vanadium dioxide film deposited by sol-gel method,} Opt. Express \textbf{22},
  %11070--11078 (2014).
%\end{thebibliography}